\documentclass[
 aip,
 amsmath,
 amssymb,
 preprint,
]{revtex4-1}

\usepackage[utf8]{inputenc}
\usepackage[T1]{fontenc}

\usepackage{graphicx}
\usepackage{dcolumn}
\usepackage{comment}
\usepackage{bm}
\usepackage{etoolbox}
\usepackage{upgreek}
\usepackage{xcolor}
\usepackage{stmaryrd}
\usepackage{subfigure}
\usepackage{array}
\usepackage{booktabs}
\usepackage{rotating}
\usepackage{float}
\usepackage{hyperref}
\usepackage{mathptmx}

\hypersetup{hidelinks}

\makeatletter
\def\@email#1#2{%
 \endgroup
 \patchcmd{\titleblock@produce}
  {\frontmatter@RRAPformat}
  {\frontmatter@RRAPformat
   {\produce@RRAP{*#1\href{mailto:#2}{#2}}}%
   \frontmatter@RRAPformat}
  {}{}
}%
\makeatother

\begin{document}
\title{Wave-branch classification of ideal MHD discontinuities in
\texorpdfstring{$Q$}{Q}-variables}
\author{Anna Krupka}
\affiliation{Centre for mathematical Plasma Astrophysics, Department of Mathematics, KU Leuven, Celestijnenlaan 200B, B-3001 Leuven, Belgium}
\email{anna.krupka@kuleuven.be}

\author{Tom Van Doorsselaere}
\affiliation{Centre for mathematical Plasma Astrophysics, Department of Mathematics, KU Leuven, Celestijnenlaan 200B, B-3001 Leuven, Belgium}

\begin{abstract}
The $Q$-variable formalism combines velocity and magnetic field into branches aligned with a selected MHD wave. For homogeneous ideal MHD, appropriate choices of the branch parameter $\alpha$ correspond to the fast, Alfv\'en, and slow characteristic families, with the background variables $\boldsymbol{Q}_0^{\pm}$ representing the propagation direction and speed of the corresponding characteristics. In structured plasmas, $\alpha$ may instead be selected from the dispersion relation of a global mode such as a kink wave. Recent work reformulated the ideal-MHD Rankine-Hugoniot conditions in $Q$-variables and established their exact equivalence with the classical conservative jump conditions. Here, we apply this formulation to the standard ideal-MHD discontinuity catalogue. We determine how the two discontinuity-frame branches $\boldsymbol{Q}^{\prime+}$ and $\boldsymbol{Q}^{\prime-}$, their sums, and their $\alpha$-normalised differences are constrained across contact, tangential, and rotational discontinuities and across the principal shock families. Material interfaces are distinguished by their normal branch content, rotational discontinuities by an Alfv\'enic tangential relation, and fast, slow, and intermediate shocks by the magnitude or orientation of the $\alpha$-normalised tangential branch difference. Perpendicular, parallel, switch-on, and switch-off shocks appear as geometrical or limiting cases. The resulting classification expresses the classical discontinuity conditions in a wave-aligned form and, through the characteristic interpretation of $\boldsymbol{Q}_0^{\pm}$, shows how ideal-MHD discontinuities constrain changes in characteristic propagation direction and speed across the interface. It therefore establishes the background interface structure required for subsequent studies of reflection, transmission, coupling, and mode conversion.
\end{abstract}
\maketitle

\section{Introduction}
\label{sec:introduction}

Discontinuities are among the most fundamental structures in magnetised
plasmas. Shocks, contact surfaces, tangential discontinuities, and
rotational discontinuities occur throughout space and astrophysical
environments, including the solar wind \citep{Bruno_Carbone_2013},
coronal mass-ejection-driven shocks \citep{Vrsnak_2008}, the Earth's
bow shock and magnetosheath \citep{Treumann_2009}, and the heliospheric
termination region
\citep{Stone_2005,Stone_2008,Baranov_1971}. Exceptionally strong
heliospheric shocks may also provide in-situ analogues of shocks in more
distant astrophysical environments \citep{Jebaraj_2024}.
Beyond the heliosphere, MHD shocks are also found in stellar and
interstellar environments, including stellar winds, accretion flows,
astrophysical jets, supernova remnants, and molecular clouds
\citep{KULSRUD_2005,Draine_1993,Goedbloed_Keppens_Poedts_2019}.
In the solar corona, tangential discontinuities and current-sheet-like
structures are also central to magnetic reconnection and to the
organisation of streamer and loop systems \citep{Priest_Forbes_2000}.
These discontinuities mediate the redistribution of mass, momentum,
magnetic flux, and energy, and therefore play a key role in the dynamics
of magnetohydrodynamic (MHD) plasmas.

The standard theoretical description of such structures is based on the
Rankine-Hugoniot (RH) conditions, obtained by applying the conservation
laws of ideal MHD across an infinitesimally thin interface
\citep{deHoffmann_Teller_1950, Jeffrey_Taniuti_2000,
Landau_Lifshitz_1984, Goedbloed_Keppens_Poedts_2019}. These conditions
provide the basis for the classical classification of MHD
discontinuities. In this framework, contact and tangential
discontinuities, rotational discontinuities, and fast, slow,
intermediate, perpendicular, parallel, switch-on, and switch-off shocks
are distinguished by their jumps in the primitive variables
\((\rho,\boldsymbol{V},p,\boldsymbol{B})\), by the presence or absence
of normal mass flux and magnetic connection, and by their relation to the
characteristic speeds of the MHD system
\citep{deHoffmann_Teller_1950, Jeffrey_Taniuti_2000,
Goedbloed_Keppens_Poedts_2019}.

The present work connects the classical RH and characteristic description
of ideal-MHD discontinuities with a wave-aligned representation. The
classical RH conditions and characteristic structure organise the
physical discontinuity families, while the $Q$-variables express their
velocity-magnetic-field structure in a basis selected through $\alpha$.
For a branch-specific choice of $\alpha$, the corresponding background
$Q$-variables are directly associated with the fast, Alfv\'en, or slow
MHD characteristics. The $Q$-variable jump relations therefore constrain
not only the wave-aligned branch content on the two sides of the
interface, but also how the propagation direction and speed of the
associated characteristics change across the discontinuity.

This connection is particularly useful for wave-interface problems. A
discontinuity supplies the nonlinear background matching conditions,
while an incident Alfv\'en, fast, slow, kink, or other disturbance
supplies a specific wave branch. Recent studies of strong heliospheric
shocks and turbulence-driven shock corrugation illustrate the importance
of understanding this interface structure
\citep{Jebaraj_2024,Jebaraj_2026}.

A wave-based description of MHD is most familiar through the Els\"asser
variables \citep{Elsasser_1950},
\[
\boldsymbol{Z}^{\pm}
=
\boldsymbol{V}
\pm
\boldsymbol{V}_{A},
\]
where $\boldsymbol{V}$ is the plasma velocity and
\[
\boldsymbol{V}_{A}
=
\frac{\boldsymbol{B}}{\sqrt{\mu\rho}}
\]
is the Alfv\'en velocity, with $\boldsymbol{B}$ the magnetic field,
$\rho$ the plasma mass density, and $\mu$ the magnetic permeability.
These variables separate oppositely propagating Alfv\'enic contributions
and have become a central description of Alfv\'enic turbulence in the
solar wind
\citep{Bruno_Carbone_2013, Dobrowolny_1980, Marsch_Tu_1989,
Tu_Marsch_Thieme_1989, Velli_Grappin_Mangeney_1989,
Zhou_Matthaeus_1989, Grappin_1990}. Their success has motivated broader
attempts to formulate MHD dynamics in terms of variables associated with
propagation directions \citep{Marsch_Mangeney_1987}.

Els\"asser variables, however, are tied to the Alfv\'en speed and
therefore do not provide a mode-independent representation of the full
MHD wave system. Magneto-acoustic disturbances involve mixed
contributions from both Els\"asser fields even in homogeneous plasma
\citep{Magyar_2019}, and this mixing persists in non-uniform
environments \citep{Ismayilli_2022}. More generally, the extension of
Els\"asser-type descriptions beyond purely Alfv\'enic dynamics is limited
by the fact that the fast, slow, and Alfv\'enic branches have different
characteristic speeds and different couplings to thermodynamic variables
\citep{Galtier_2023, Schekochihin_2009}. This motivates the use of more
general wave-aligned variables.

The $Q$-variable formalism introduced by \citet{VanDoorsselaere_2024}
generalises the Els\"asser representation by defining
\[
    \boldsymbol{Q}^{\pm}
    =
    \boldsymbol{V}
    \pm
    \alpha\boldsymbol{B},
\]
where the branch parameter \(\alpha\) is selected from the phase relation
of the wave under consideration. For a homogeneous background state, we
write
\[
    \boldsymbol{Q}_{0}^{\pm}
    =
    \boldsymbol{V}_{0}
    \pm
    \alpha_{0}\boldsymbol{B}_{0}.
\]
Here, the subscript \(0\) denotes the equilibrium or background state
about which a linear wave perturbation may be introduced.
For the branch-specific choice of \(\alpha_0\) corresponding to a fast,
Alfv\'en, or slow MHD wave, the background \(Q\)-variables may be
identified with the corresponding MHD characteristics. The characteristic curves satisfy
\[
    Q_{0,m,n}^{\pm}\equiv\lambda_{m}^{\pm}
    =
    V_{0n}\pm c_m,
    \qquad
    m\in\{f,A,s\},
\]
where \(c_f\), \(c_A\), and \(c_s\) are the fast, Alfv\'en, and slow
characteristic speeds, respectively. Thus,
\[
    \lambda_f^\pm = Q_{0,f,n}^\pm,
    \qquad
    \lambda_A^\pm = Q_{0,A,n}^\pm,
    \qquad
    \lambda_s^\pm = Q_{0,s,n}^\pm,
\]
so that the three branch choices and their two propagation directions
represent the six propagating characteristics of ideal MHD. The
background vectors \(\boldsymbol{Q}_{0,m}^{\pm}\) therefore provide a
direct wave-aligned representation of the corresponding characteristic
propagation, while their projection onto \(\boldsymbol{n}\) gives the
characteristic speed \(\lambda_m^\pm\).

For the Alfv\'enic branch,
\(\alpha=1/\sqrt{\mu\rho}\), and the $Q$-variables reduce to the
Els\"asser variables. For homogeneous fast and slow waves, \(\alpha\) is
obtained from the corresponding magneto-acoustic dispersion relation. In
structured plasmas, it may instead be associated with the phase speed of
a global mode, including kink waves \citep{VanDoorsselaere_2024}. The
$Q$-formalism is therefore broader than a transformation to the local
ideal-MHD characteristic variables: it provides a common wave-aligned
representation for both characteristic waves and structured-plasma
modes.

The RH conditions were recently reformulated in terms of the $Q$-variables
by \citet{krupka2026rankine}. In that work, the full ideal-MHD jump
conditions were derived directly in terms of
\(\boldsymbol{Q}^{\pm}\) and were shown to be exactly equivalent to the
classical RH conditions for mass, momentum, magnetic flux, and energy.
That result established the $Q$-variable RH system as a conservative
formulation of discontinuities. The next step is to use this formulation
as a way to extract the wave-branch structure hidden inside the classical
discontinuity catalogue and to interpret the resulting branch changes in
terms of the corresponding MHD characteristics.

The purpose of the present work is to construct this wave-aligned
classification. For each classical discontinuity family, we impose its
defining physical conditions on the $Q$-variable RH system and determine
how the discontinuity-frame branches
$\boldsymbol{Q}^{\prime+}$ and $\boldsymbol{Q}^{\prime-}$, their sums,
and their $\alpha$-normalised differences are related across the
interface. The resulting map identifies which branch combinations are
continuous or coupled, which become equal or vanish, and how their
magnitudes or orientations differ between the two sides. Through the
identification of the background $Q$-variables with the corresponding
MHD characteristics, these relations also show how each discontinuity
constrains changes in characteristic propagation direction and speed.

This classification is also a necessary step toward a broader
$Q$-variable description of wave-discontinuity interactions. Before
asking how a specific Alfv\'en, fast, slow, or kink disturbance interacts
with a discontinuity, one must first identify the $Q$-space signature of
the discontinuity itself. The present paper provides this branch-level
map.

This paper is organised as follows.
Section~\ref{sec:Qframework} summarises the $Q$-variable RH framework
used throughout the paper.
Section~\ref{sec:noncompressive} derives the $Q$-space signatures of
contact, tangential, and rotational discontinuities.
Section~\ref{sec:magneto-acoustic_shocks} analyses magneto-acoustic
shocks and the fast, slow, and intermediate shock families.
Section~\ref{sec:geometrical_limiting_shocks} considers perpendicular,
parallel, switch-on, and switch-off shocks as geometrical and limiting
cases.
Section~\ref{sec:classification_table} collects the resulting signatures
in a diagnostic classification table.
Section~\ref{sec:discussion} discusses their physical interpretation,
diagnostic value, and limitations.
Finally, Section~\ref{sec:conclusions} summarises the main conclusions.
\section{\texorpdfstring{$Q$}{Q}-variable Rankine-Hugoniot framework}
\label{sec:Qframework}

We briefly summarise the $Q$-variable Rankine-Hugoniot ($Q$-RH) framework used throughout this paper. The full derivation of the $Q$-variable jump conditions, including the energy condition, and their exact equivalence with the classical ideal-MHD RH conditions were presented in \citet{krupka2026rankine}. Here, we use that conservative system as the starting point for extracting the branch-level structure of ideal-MHD discontinuities.

The $Q$-variable formalism employs two wave-aligned variables,
\begin{equation}
    \boldsymbol{Q}^{\pm}
    =
    \boldsymbol{V}
    \pm
    \alpha \boldsymbol{B}.
    \label{eq:Q_definition}
\end{equation}
We consider a planar discontinuity separating an upstream state, labelled by subscript~1, from a downstream state, labelled by subscript~2. The jump of any quantity \(f\) is defined as
\begin{equation}
    \llbracket f \rrbracket
    \equiv
    f_1-f_2 .
    \label{eq:jump_definition}
\end{equation}
The unit normal \(\boldsymbol{n}\) is taken to point from the downstream region toward the upstream region. If the discontinuity propagates with normal speed \(u\), the shock-frame $Q$-variables are
\begin{equation}
    \boldsymbol{Q}^{\prime\pm}
    =
    \boldsymbol{Q}^{\pm}
    -
    u\boldsymbol{n}.
    \label{eq:shock_frame_Q}
\end{equation}
Their normal and tangential components are
\begin{equation}
    Q_n^{\prime\pm}
    =
    \boldsymbol{Q}^{\prime\pm}\cdot\boldsymbol{n},
    \qquad
    \boldsymbol{Q}_t^{\prime\pm}
    =
    \boldsymbol{Q}^{\prime\pm}
    -
    Q_n^{\prime\pm}\boldsymbol{n}.
    \label{eq:normal_tangential_Q}
\end{equation}
Since the shock-frame transformation affects only the normal velocity, the tangential components satisfy \(\boldsymbol{Q}_t^{\prime\pm}=\boldsymbol{Q}_t^{\pm}\). Equivalently, introducing the shock-frame velocity
$\boldsymbol{V}'=\boldsymbol{V}-u\boldsymbol{n}$, the primitive velocity
and magnetic field are recovered from the branch sum and difference as
\begin{equation}
    \boldsymbol{V}'
    =
    \frac{1}{2}
    \left(
        \boldsymbol{Q}^{\prime+}
        +
        \boldsymbol{Q}^{\prime-}
    \right),
    \qquad
    \boldsymbol{B}
    =
    \frac{1}{2\alpha}
    \left(
        \boldsymbol{Q}^{\prime+}
        -
        \boldsymbol{Q}^{\prime-}
    \right).
    \label{eq:Q_inverse_relations}
\end{equation}
The shock-frame transformation therefore cancels from the branch
difference,
\begin{equation}
    \boldsymbol{Q}^{\prime+}
    -
    \boldsymbol{Q}^{\prime-}
    =
    2\alpha\boldsymbol{B}.
    \label{eq:Q_branch_difference}
\end{equation}
Thus, the branch sum represents twice the shock-frame velocity, whereas
the $\alpha$-normalised branch difference represents twice the magnetic
field.

The conservative $Q$-RH conditions may then be written as \citep{krupka2026rankine}
\begin{align}
    &\bigg\llbracket
        \frac{Q^{\prime+}_n - Q^{\prime-}_n}{\alpha}
    \bigg\rrbracket
    = 0,
    \label{eq:QRH_normal_flux}
    \\[4pt]
    &\bigg\llbracket
        \rho
        \left(
            Q^{\prime+}_n + Q^{\prime-}_n
        \right)
    \bigg\rrbracket
    = 0,
    \label{eq:QRH_mass}
    \\[4pt]
    &\bigg\llbracket
        \frac{
            Q^{\prime-}_n \boldsymbol{Q}^{\prime+}_t
            -
            Q^{\prime+}_n \boldsymbol{Q}^{\prime-}_t
        }{\alpha}
    \bigg\rrbracket
    = 0,
    \label{eq:QRH_tangential_flux}
    \\[6pt]
    &\bigg\llbracket
        \rho
        \left(
            Q^{\prime+}_n+Q^{\prime-}_n
        \right)^2
        +
        4p
        +
        \frac{1}{2\mu\alpha^2}
        \left(
            \boldsymbol{Q}_t^{\prime+}
            -
            \boldsymbol{Q}_t^{\prime-}
        \right)^2
    \bigg\rrbracket
    = 0,
    \label{eq:QRH_normal_momentum}
    \\[6pt]
    &\bigg\llbracket
        \rho
        \left(
            Q^{\prime+}_n+Q^{\prime-}_n
        \right)
        \left(
            \boldsymbol{Q}^{\prime+}_t+\boldsymbol{Q}^{\prime-}_t
        \right)
        -
        \frac{1}{\mu\alpha^2}
        \left(
            Q^{\prime+}_n - Q^{\prime-}_n
        \right)
        \left(
            \boldsymbol{Q}^{\prime+}_t - \boldsymbol{Q}^{\prime-}_t
        \right)
    \bigg\rrbracket
    = 0.
    \label{eq:QRH_tangential_momentum}
\end{align}
These equations express, respectively, continuity of normal magnetic flux, mass conservation, tangential magnetic-flux conservation, normal momentum conservation, and tangential momentum conservation in $Q$-variables. The full conservative $Q$-RH system also contains the energy condition
derived in \citet{krupka2026rankine}. The branch relations used in the
present classification follow from the mass, magnetic-flux, and momentum
conditions written above, while the energy condition completes the
thermodynamic closure of a full compressive-shock solution.

Equations~\eqref{eq:QRH_normal_flux}-\eqref{eq:QRH_tangential_momentum}
provide the algebraic basis for the following derivations. For each
discontinuity family, we impose its defining classical conditions on this
system and identify the resulting constraints on the discontinuity-frame
$Q$-branches.

\section{Non-compressive discontinuities}
\label{sec:noncompressive}

We first consider contact and tangential discontinuities, which are
material interfaces, before turning to rotational discontinuities. For a
material interface, the normal velocity vanishes in the discontinuity
frame,
\begin{equation}
    V_n'=0.
\end{equation}
Using Eq.~\eqref{eq:Q_inverse_relations}, this gives
\begin{equation}
    Q_n^{\prime+}
    +
    Q_n^{\prime-}
    =
    0,
    \label{eq:no_mass_flux_Q}
\end{equation}
or equivalently
\begin{equation}
    Q_n^{\prime+}
    =
    -
    Q_n^{\prime-}.
    \label{eq:no_mass_flux_Q_pm}
\end{equation}
These relations are understood to hold on both sides of the interface.
The $Q$-variable mass condition, Eq.~\eqref{eq:QRH_mass}, is therefore
automatically satisfied, independently of whether the density changes
across the interface.

The normal magnetic field is represented in $Q$-variables by
\begin{equation}
    B_n
    =
    \frac{
        Q_n^{\prime+}
        -
        Q_n^{\prime-}
    }{
        2\alpha
    }.
    \label{eq:Bn_Q_noncompressive}
\end{equation}
Under the no-mass-flux constraint~\eqref{eq:no_mass_flux_Q}, this becomes
\begin{equation}
    B_n
    =
    \frac{Q_n^{\prime+}}{\alpha}
    =
    -
    \frac{Q_n^{\prime-}}{\alpha}.
    \label{eq:Bn_Q_pm_noncompressive}
\end{equation}

\subsection{Contact discontinuities}
\label{sec:contact_discontinuity}

A contact discontinuity is a material interface with finite normal magnetic
connection \cite{Jeffrey_Taniuti_2000,Goedbloed_Keppens_Poedts_2019}.
Using Eq.~\eqref{eq:Bn_Q_pm_noncompressive}, this gives
\begin{equation}
    \frac{Q_n^{\prime\pm}}{\alpha}
    \neq
    0.
\end{equation}

The normal magnetic-flux condition gives
\begin{equation}
    \bigg\llbracket
        \frac{
            Q_n^{\prime+}
            -
            Q_n^{\prime-}
        }{\alpha}
    \bigg\rrbracket
    =
    0.
    \label{eq:contact_normal_flux_difference}
\end{equation}
Using Eq.~\eqref{eq:no_mass_flux_Q_pm}, this may be written as
\begin{equation}
    \bigg\llbracket
        \frac{
            Q_n^{\prime\pm}
        }{\alpha}
    \bigg\rrbracket
    =
    0.
    \label{eq:contact_normal_flux_Q}
\end{equation}
Thus, the finite $\alpha$-normalised normal branch components are
continuous across a contact discontinuity.

Since \(Q_n^{\prime\pm}/\alpha\neq0\), the tangential magnetic-flux condition,
Eq.~\eqref{eq:QRH_tangential_flux}, and the tangential momentum condition,
Eq.~\eqref{eq:QRH_tangential_momentum}, imply continuity of the tangential
branch sum and of the normalised tangential branch difference:
\begin{align}
    &\bigg\llbracket
        \boldsymbol{Q}_t^{\prime+}
        +
        \boldsymbol{Q}_t^{\prime-}
    \bigg\rrbracket
    =
    0,
    \label{eq:contact_tangential_sum_Q}
    \\[4pt]
    &\bigg\llbracket
        \frac{
            \boldsymbol{Q}_t^{\prime+}
            -
            \boldsymbol{Q}_t^{\prime-}
        }{\alpha}
    \bigg\rrbracket
    =
    0.
    \label{eq:contact_tangential_separation_Q}
\end{align}
The normal momentum condition, Eq.~\eqref{eq:QRH_normal_momentum}, then
reduces to gas-pressure balance,
\begin{equation}
    \llbracket p \rrbracket
    =
    0.
    \label{eq:contact_pressure_Q}
\end{equation}
The density is not fixed by the contact-discontinuity jump conditions and
may therefore differ on the two sides of the interface.

Collecting these results, a contact discontinuity in $Q$-variables is described by
\begin{align}
    &Q_n^{\prime+}
    =
    -
    Q_n^{\prime-},
    \qquad
    \frac{Q_n^{\prime\pm}}{\alpha}
    \neq
    0,
    \qquad
    \left\llbracket
        \frac{Q_n^{\prime\pm}}{\alpha}
    \right\rrbracket
    =
    0,
    \nonumber\\[3pt]
    &\left\llbracket
        \boldsymbol Q_t^{\prime+}
        +
        \boldsymbol Q_t^{\prime-}
    \right\rrbracket
    =
    0,
    \qquad
    \left\llbracket
        \frac{
            \boldsymbol Q_t^{\prime+}
            -
            \boldsymbol Q_t^{\prime-}
        }{\alpha}
    \right\rrbracket
    =
    0,
    \qquad
    \llbracket p\rrbracket
    =
    0.
    \label{eq:contact_summary_Q}
\end{align}
Thus, a contact discontinuity has finite, opposite normal branch
components whose $\alpha$-normalised values are continuous across the
interface. The tangential branch sum and the $\alpha$-normalised
tangential branch difference are also continuous, while the density may
differ between the two sides.

In terms of the background characteristics introduced in
Section~\ref{sec:introduction}, the finite normal branch components imply
that the corresponding characteristics retain a finite component across
the interface. The velocity and magnetic field are continuous across the
contact discontinuity, but a density jump may change the branch-specific
parameter $\alpha_m$ and hence the propagation speed and, in general, the
orientation of the associated characteristic. A contact discontinuity can
therefore modify the characteristic through the change of background
plasma state even though the velocity and magnetic field themselves remain
continuous.

\subsection{Tangential discontinuities}
\label{sec:tangential_discontinuity}

A tangential discontinuity is a material interface with no normal magnetic
connection \cite{Jeffrey_Taniuti_2000,Goedbloed_Keppens_Poedts_2019}.
In $Q$-space, this means that the no-mass-flux condition is accompanied by
vanishing of the normal branch components. Combining
Eq.~\eqref{eq:no_mass_flux_Q} with the absence of normal magnetic
connection gives
\begin{equation}
    Q_n^{\prime+}
    =
    Q_n^{\prime-}
    =
    0.
    \label{eq:tangential_zero_normal_Q}
\end{equation}
Thus, both normal branch components vanish in the discontinuity frame.

With Eq.~\eqref{eq:tangential_zero_normal_Q}, the mass condition,
Eq.~\eqref{eq:QRH_mass}, the normal magnetic-flux condition,
Eq.~\eqref{eq:QRH_normal_flux}, the tangential magnetic-flux condition,
Eq.~\eqref{eq:QRH_tangential_flux}, and the tangential momentum
condition, Eq.~\eqref{eq:QRH_tangential_momentum}, are automatically
satisfied. Neither the tangential branch sum nor the $\alpha$-normalised
tangential branch difference is required to be continuous across a
tangential discontinuity. Unlike a contact discontinuity, a tangential
discontinuity may consequently support jumps in its tangential branch
structure.

The only remaining non-trivial condition is the normal momentum balance.
In $Q$-variables this becomes
\begin{equation}
    \bigg\llbracket
        4p
        +
        \frac{1}{2\mu\alpha^2}
        \left|
            \boldsymbol{Q}_t^{\prime+}
            -
            \boldsymbol{Q}_t^{\prime-}
        \right|^2
    \bigg\rrbracket
    =
    0.
    \label{eq:TD_total_pressure_Q}
\end{equation}
This is the $Q$-variable form of total pressure balance. The gas pressure
and tangential branch difference may each change, but their combined
contribution must remain balanced across the interface. The density is
also not fixed by the tangential-discontinuity jump conditions.

Collecting these results, a tangential discontinuity in $Q$-variables is
described by
\begin{align}
    &Q_n^{\prime+}
    =
    Q_n^{\prime-}
    =
    0,
    \nonumber\\[3pt]
    &\left\llbracket
        4p
        +
        \frac{1}{2\mu\alpha^2}
        \left|
            \boldsymbol{Q}_t^{\prime+}
            -
            \boldsymbol{Q}_t^{\prime-}
        \right|^2
    \right\rrbracket
    =
    0.
    \label{eq:tangential_summary_Q}
\end{align}

The tangential branch sum and the $\alpha$-normalised tangential branch
difference may therefore differ between the two sides, subject to the
total-pressure balance in Eq.~\eqref{eq:TD_total_pressure_Q}. A
tangential discontinuity consequently has vanishing normal branch
components, while its wave-aligned structure is entirely tangential.

In the characteristic interpretation, the vanishing normal branch
components mean that the corresponding background characteristics have
no normal propagation across the discontinuity in its rest frame. Their
structure is therefore confined to the tangential direction. Since the
tangential branch components need not be continuous, the propagation
speed and direction of the corresponding tangential characteristics may
differ on the two sides of the interface, while no characteristic
component is transmitted normally through the discontinuity.

An idealised coronal-loop boundary provides a familiar example. In a
piecewise-uniform equilibrium, the magnetic field is tangential to the
loop boundary and the normal plasma velocity vanishes, while the density
and tangential magnetic field may differ between the loop and its
surroundings subject to total-pressure balance
\citep{Edwin_Roberts_1983}. A finite transition layer gives the
corresponding continuously structured configuration.

\subsection{Rotational discontinuities}
\label{sec:rotational_discontinuity}

We now consider rotational discontinuities, which are non-compressive
finite-flux discontinuities with finite normal velocity and finite normal
magnetic field
\citep{deHoffmann_Teller_1950,Goedbloed_Keppens_Poedts_2019}. In
$Q$-variables, this class is characterised by
\begin{equation}
    \llbracket \rho \rrbracket
    =
    0,
    \qquad
    \llbracket p \rrbracket
    =
    0,
    \qquad
    Q_n^{\prime+}
    +
    Q_n^{\prime-}
    \neq
    0,
    \qquad
    Q_n^{\prime+}
    \neq
    Q_n^{\prime-}.
    \label{eq:RD_basic_conditions_Q}
\end{equation}
Using Eq.~\eqref{eq:Q_inverse_relations}, the third condition is
equivalent to \(V_n'\neq0\) and represents finite normal mass flux,
whereas the fourth is equivalent to \(B_n\neq0\) and represents finite
normal magnetic connection.

The $Q$-variable mass conservation condition is
\begin{equation}
    \bigg\llbracket
        \rho
        \left(
            Q_n^{\prime+}
            +
            Q_n^{\prime-}
        \right)
    \bigg\rrbracket
    =
    0.
    \label{eq:RD_mass_Q}
\end{equation}
Since the density is continuous across a rotational discontinuity, this gives
\begin{equation}
    \bigg\llbracket
        Q_n^{\prime+}
        +
        Q_n^{\prime-}
    \bigg\rrbracket
    =
    0.
    \label{eq:RD_normal_sum_Q}
\end{equation}
Thus, the normal branch sum is finite and continuous across the
discontinuity.

The normal magnetic-flux condition is
\begin{equation}
    \bigg\llbracket
        \frac{
            Q_n^{\prime+}
            -
            Q_n^{\prime-}
        }{\alpha}
    \bigg\rrbracket
    =
    0.
    \label{eq:RD_normal_separation_Q}
\end{equation}
Thus, the $\alpha$-normalised normal branch difference is also
continuous across the discontinuity.

The characteristic Alfv\'enic condition follows from the existence of a
non-trivial tangential jump. We derive it directly from the $Q$-variable
tangential magnetic-flux and tangential momentum conditions. The
tangential magnetic-flux condition is
\begin{equation}
    \bigg\llbracket
        \frac{
            Q_n^{\prime-}\boldsymbol{Q}_t^{\prime+}
            -
            Q_n^{\prime+}\boldsymbol{Q}_t^{\prime-}
        }{\alpha}
    \bigg\rrbracket
    =
    0.
    \label{eq:RD_tangential_flux_full_Q}
\end{equation}
Using Eqs.~\eqref{eq:RD_normal_sum_Q}
and~\eqref{eq:RD_normal_separation_Q}, this can be written as
\begin{equation}
    \left(
        Q_n^{\prime+}
        +
        Q_n^{\prime-}
    \right)
    \bigg\llbracket
        \frac{
            \boldsymbol{Q}_t^{\prime+}
            -
            \boldsymbol{Q}_t^{\prime-}
        }{\alpha}
    \bigg\rrbracket
    -
    \frac{
        Q_n^{\prime+}
        -
        Q_n^{\prime-}
    }{\alpha}
    \bigg\llbracket
        \boldsymbol{Q}_t^{\prime+}
        +
        \boldsymbol{Q}_t^{\prime-}
    \bigg\rrbracket
    =
    0.
    \label{eq:RD_tangential_flux_reduced_Q}
\end{equation}
Similarly, the tangential momentum condition,
\begin{equation}
    \bigg\llbracket
        \rho
        \left(
            Q_n^{\prime+}
            +
            Q_n^{\prime-}
        \right)
        \left(
            \boldsymbol{Q}_t^{\prime+}
            +
            \boldsymbol{Q}_t^{\prime-}
        \right)
        -
        \frac{1}{\mu\alpha^2}
        \left(
            Q_n^{\prime+}
            -
            Q_n^{\prime-}
        \right)
        \left(
            \boldsymbol{Q}_t^{\prime+}
            -
            \boldsymbol{Q}_t^{\prime-}
        \right)
    \bigg\rrbracket
    =
    0,
    \label{eq:RD_tangential_momentum_full_Q}
\end{equation}
reduces to
\begin{equation}
    \rho
    \left(
        Q_n^{\prime+}
        +
        Q_n^{\prime-}
    \right)
    \bigg\llbracket
        \boldsymbol{Q}_t^{\prime+}
        +
        \boldsymbol{Q}_t^{\prime-}
    \bigg\rrbracket
    -
    \frac{1}{\mu}
    \frac{
        Q_n^{\prime+}
        -
        Q_n^{\prime-}
    }{\alpha}
    \bigg\llbracket
        \frac{
            \boldsymbol{Q}_t^{\prime+}
            -
            \boldsymbol{Q}_t^{\prime-}
        }{\alpha}
    \bigg\rrbracket
    =
    0.
    \label{eq:RD_tangential_momentum_reduced_Q}
\end{equation}

Equations~\eqref{eq:RD_tangential_flux_reduced_Q}
and~\eqref{eq:RD_tangential_momentum_reduced_Q} are two homogeneous
relations for the jumps in the tangential branch sum and in the
normalised tangential branch difference. Since a rotational
discontinuity has a non-trivial tangential jump, a non-zero solution
requires
\begin{equation}
    \rho
    \left(
        Q_n^{\prime+}
        +
        Q_n^{\prime-}
    \right)^2
    =
    \frac{1}{\mu}
    \left(
        \frac{
            Q_n^{\prime+}
            -
            Q_n^{\prime-}
        }{\alpha}
    \right)^2 .
    \label{eq:RD_Alfven_condition_Q}
\end{equation}
Equivalently, defining the normal Alfv\'en Mach number in $Q$-form as
\begin{equation}
    M^2
    \equiv
    \frac{
        \left(
            Q_n^{\prime+}
            +
            Q_n^{\prime-}
        \right)^2
        \alpha^2\mu\rho
    }{
        \left(
            Q_n^{\prime+}
            -
            Q_n^{\prime-}
        \right)^2
    },
    \label{eq:RD_Mach_Q}
\end{equation}
one obtains
\begin{equation}
    M^2
    =
    1.
    \label{eq:RD_Mach_one_Q}
\end{equation}
Thus, the $Q$-variable tangential flux and tangential momentum balances
place a rotational discontinuity on the normal Alfv\'enic
characteristic, with equality between the normal flow speed and the
normal Alfv\'en speed. For the Alfv\'enic choice
\(\alpha_A=1/\sqrt{\mu\rho}\), this means that one of the two Alfv\'enic
characteristics has zero normal propagation speed in the discontinuity
frame: the rotational discontinuity itself propagates along that
characteristic.

The same reduced equations determine how the tangential branch structure
changes across the discontinuity. From
Eq.~\eqref{eq:RD_Alfven_condition_Q}, one has
\begin{equation}
    \frac{
        Q_n^{\prime+}
        -
        Q_n^{\prime-}
    }{\alpha}
    =
    \pm
    \sqrt{\mu\rho}
    \left(
        Q_n^{\prime+}
        +
        Q_n^{\prime-}
    \right).
    \label{eq:RD_normal_branch_relation_Q}
\end{equation}
Substituting this relation into
Eq.~\eqref{eq:RD_tangential_flux_reduced_Q} gives
\begin{equation}
    \bigg\llbracket
        \boldsymbol{Q}_t^{\prime+}
        +
        \boldsymbol{Q}_t^{\prime-}
    \bigg\rrbracket
    =
    \pm
    \frac{1}{\sqrt{\mu\rho}}
    \bigg\llbracket
        \frac{
            \boldsymbol{Q}_t^{\prime+}
            -
            \boldsymbol{Q}_t^{\prime-}
        }{\alpha}
    \bigg\rrbracket .
    \label{eq:RD_tangential_relation_Q}
\end{equation}
This relation ties the jump in the tangential branch sum to the jump in
the $\alpha$-normalised tangential branch difference. The two
combinations therefore cannot change independently but are linked by an
Alfv\'enic relation. This is the $Q$-space expression of tangential
rotation across a rotational discontinuity.

The remaining rotational property follows from the normal momentum
condition,
\begin{equation}
    \bigg\llbracket
        \rho
        \left(
            Q_n^{\prime+}
            +
            Q_n^{\prime-}
        \right)^2
        +
        4p
        +
        \frac{1}{2\mu\alpha^2}
        \left|
            \boldsymbol{Q}_t^{\prime+}
            -
            \boldsymbol{Q}_t^{\prime-}
        \right|^2
    \bigg\rrbracket
    =
    0.
    \label{eq:RD_normal_momentum_Q}
\end{equation}
Using Eqs.~\eqref{eq:RD_basic_conditions_Q}
and~\eqref{eq:RD_normal_sum_Q}, this reduces to
\begin{equation}
    \bigg\llbracket
        \frac{
            \left|
                \boldsymbol{Q}_t^{\prime+}
                -
                \boldsymbol{Q}_t^{\prime-}
            \right|^2
        }{\alpha^2}
    \bigg\rrbracket
    =
    0.
    \label{eq:RD_tangential_invariant_Q}
\end{equation}
Thus, the magnitude of the $\alpha$-normalised tangential branch
difference is continuous across the discontinuity. Its direction may
change, but its magnitude cannot.

Collecting the results, a rotational discontinuity is described in
$Q$-variables by
\begin{align}
    &\llbracket \rho \rrbracket
    =
    0,
    \qquad
    \llbracket p \rrbracket
    =
    0,
    \nonumber\\[4pt]
    &\bigg\llbracket
        Q_n^{\prime+}
        +
        Q_n^{\prime-}
    \bigg\rrbracket
    =
    0,
    \nonumber\\[4pt]
    &\bigg\llbracket
        \frac{
            Q_n^{\prime+}
            -
            Q_n^{\prime-}
        }{\alpha}
    \bigg\rrbracket
    =
    0,
    \nonumber\\[4pt]
    &M^2
    =
    1,
    \nonumber\\[4pt]
    &\bigg\llbracket
        \boldsymbol{Q}_t^{\prime+}
        +
        \boldsymbol{Q}_t^{\prime-}
    \bigg\rrbracket
    =
    \pm
    \frac{1}{\sqrt{\mu\rho}}
    \bigg\llbracket
        \frac{
            \boldsymbol{Q}_t^{\prime+}
            -
            \boldsymbol{Q}_t^{\prime-}
        }{\alpha}
    \bigg\rrbracket,
    \nonumber\\[4pt]
    &\bigg\llbracket
        \frac{
            \left(
                \boldsymbol{Q}_t^{\prime+}
                -
                \boldsymbol{Q}_t^{\prime-}
            \right)^2
        }{\alpha^2}
    \bigg\rrbracket
    =
    0.
    \label{eq:RD_summary_Q}
\end{align}
A rotational discontinuity therefore has finite and continuous normal
branch sum and $\alpha$-normalised normal branch difference, with the
normal Alfv\'en Mach number satisfying \(M^2=1\). Its tangential branch
sum and $\alpha$-normalised tangential branch difference are linked by
Eq.~\eqref{eq:RD_tangential_relation_Q}, allowing the tangential
structure to rotate while its normalised magnitude remains continuous.

In characteristic terms, the rotational discontinuity is therefore tied
directly to an Alfv\'enic characteristic. The normal characteristic
condition is preserved across the interface, while the tangential
$Q$-structure changes orientation. The associated Alfv\'enic
characteristic can consequently be bent across the discontinuity without
leaving the Alfv\'enic characteristic condition; its direction changes
while the corresponding normalised tangential magnitude is preserved.

\section{Magneto-acoustic shocks}
\label{sec:magneto-acoustic_shocks}

We now turn to magneto-acoustic shocks, which are compressive
discontinuities with finite normal mass flux. Their basic conditions are
\begin{equation}
    \llbracket \rho \rrbracket
    \neq
    0,
    \qquad
    Q_n^{\prime+}
    +
    Q_n^{\prime-}
    \neq
    0.
    \label{eq:shock_basic_conditions}
\end{equation}
The second condition is understood to hold on both sides of the shock.

The $Q$-variable mass conservation condition,
Eq.~\eqref{eq:QRH_mass}, gives
\begin{equation}
    \bigg\llbracket
        \rho
        \left(
            Q_n^{\prime+}
            +
            Q_n^{\prime-}
        \right)
    \bigg\rrbracket
    =
    0.
    \label{eq:mass_flux_Q_jump}
\end{equation}
Since the normal branch sum is non-zero while the density changes,
Eq.~\eqref{eq:mass_flux_Q_jump} also implies
\begin{equation}
    \bigg\llbracket
        Q_n^{\prime+}
        +
        Q_n^{\prime-}
    \bigg\rrbracket
    \neq
    0.
    \label{eq:shock_normal_sum_jump}
\end{equation}
Thus, a compressive shock has a continuous density-weighted normal branch
sum, while the unweighted normal branch sum changes across the shock. This
change is the direct $Q$-space consequence of compression.

The normal branch difference satisfies the $Q$-variable normal
magnetic-flux condition, Eq.~\eqref{eq:QRH_normal_flux},
\begin{equation}
    \bigg\llbracket
        \frac{
            Q_n^{\prime+}
            -
            Q_n^{\prime-}
        }{\alpha}
    \bigg\rrbracket
    =
    0.
    \label{eq:shock_normal_flux_Q}
\end{equation}
Therefore, across a shock, the normal branch sum changes through
compression, whereas the $\alpha$-normalised normal branch difference
remains continuous. Together with
Eq.~\eqref{eq:mass_flux_Q_jump}, these relations provide the basic
normal-component signature of a compressive discontinuity in $Q$-space.

In terms of the background characteristics introduced in
Section~\ref{sec:introduction}, the compression changes the plasma state
and therefore changes the characteristic propagation on the two sides of
the shock. The $Q$-RH relations therefore
provide the matching conditions that determine how the characteristic
speed and direction change across the shock.

To classify the shock families, we combine the tangential magnetic-flux
condition, Eq.~\eqref{eq:QRH_tangential_flux}, and the tangential
momentum condition, Eq.~\eqref{eq:QRH_tangential_momentum}. This is the
shock analogue of the reduction used for rotational discontinuities in
Sec.~\ref{sec:rotational_discontinuity}, but here we do not impose
$\llbracket\rho\rrbracket=0$.

We first consider the generic case with finite normal magnetic field,
$B_n\neq0$, or equivalently
$Q_n^{\prime+}-Q_n^{\prime-}\neq0$. This allows the normal branch
difference to be divided out in the following reduction. The
perpendicular limit $B_n=0$ is treated separately below.

The tangential magnetic-flux condition can be written in terms of the
tangential branch sum and the normalised tangential branch difference as
\begin{equation}
    \bigg\llbracket
        \boldsymbol{Q}_t^{\prime+}
        +
        \boldsymbol{Q}_t^{\prime-}
    \bigg\rrbracket
    =
    \frac{\alpha}{
        Q_n^{\prime+}
        -
        Q_n^{\prime-}
    }
    \bigg\llbracket
        \left(
            Q_n^{\prime+}
            +
            Q_n^{\prime-}
        \right)
        \left(
            \frac{
                \boldsymbol{Q}_t^{\prime+}
                -
                \boldsymbol{Q}_t^{\prime-}
            }{\alpha}
        \right)
    \bigg\rrbracket .
    \label{eq:shock_tangential_flux_reduced_Q}
\end{equation}
Similarly, the tangential momentum condition gives
\begin{equation}
    \bigg\llbracket
        \frac{
            \boldsymbol{Q}_t^{\prime+}
            -
            \boldsymbol{Q}_t^{\prime-}
        }{\alpha}
    \bigg\rrbracket
    =
    \frac{
        \alpha\mu\rho
        \left(
            Q_n^{\prime+}
            +
            Q_n^{\prime-}
        \right)
    }{
        Q_n^{\prime+}
        -
        Q_n^{\prime-}
    }
    \bigg\llbracket
        \boldsymbol{Q}_t^{\prime+}
        +
        \boldsymbol{Q}_t^{\prime-}
    \bigg\rrbracket .
    \label{eq:shock_tangential_momentum_reduced_Q}
\end{equation}
Substituting Eq.~\eqref{eq:shock_tangential_flux_reduced_Q} into
Eq.~\eqref{eq:shock_tangential_momentum_reduced_Q}, and using the
conserved mass flux and normal magnetic flux, gives
\begin{equation}
    \bigg\llbracket
        \frac{
            \left(
                Q_n^{\prime+}
                +
                Q_n^{\prime-}
            \right)^2
            \alpha^2\mu\rho
        }{
            \left(
                Q_n^{\prime+}
                -
                Q_n^{\prime-}
            \right)^2
        }
        \left(
            \frac{
                \boldsymbol{Q}_t^{\prime+}
                -
                \boldsymbol{Q}_t^{\prime-}
            }{\alpha}
        \right)
    \bigg\rrbracket
    =
    \bigg\llbracket
        \frac{
            \boldsymbol{Q}_t^{\prime+}
            -
            \boldsymbol{Q}_t^{\prime-}
        }{\alpha}
    \bigg\rrbracket .
    \label{eq:shock_tangential_relation_Q}
\end{equation}

Using the normal Alfv\'en Mach number,
\begin{equation}
    M^2
    \equiv
    \frac{
        \left(
            Q_n^{\prime+}
            +
            Q_n^{\prime-}
        \right)^2
        \alpha^2\mu\rho
    }{
        \left(
            Q_n^{\prime+}
            -
            Q_n^{\prime-}
        \right)^2
    },
    \label{eq:shock_Mach_Q}
\end{equation}
Eq.~\eqref{eq:shock_tangential_relation_Q} becomes
\begin{equation}
    \bigg\llbracket
        \left(
            M^2
            -
            1
        \right)
        \frac{
            \boldsymbol{Q}_t^{\prime+}
            -
            \boldsymbol{Q}_t^{\prime-}
        }{\alpha}
    \bigg\rrbracket
    =
    0.
    \label{eq:shock_tangential_invariant_Q}
\end{equation}
Equation~\eqref{eq:shock_tangential_invariant_Q} is the central
$Q$-space relation for magneto-acoustic shocks. It connects the
$\alpha$-normalised tangential branch difference to the local normal
Alfv\'en Mach number. Since the density changes across a compressive
shock, the upstream and downstream values \(M_1^2\) and \(M_2^2\)
generally differ. Their classical characteristic ordering determines
whether the tangential branch difference increases in magnitude, decreases
in magnitude, or reverses orientation.

We first apply this invariant to the fast, slow, and intermediate
characteristic families and then consider perpendicular, parallel,
switch-on, and switch-off shocks as geometrical or limiting cases.

\subsection{Characteristic shock families}
\label{sec:characteristic_shock_families}

The fast, slow, and intermediate shock families are defined by the
ordering of the upstream and downstream states relative to the fast,
Alfv\'en, and slow characteristic speeds
\citep{Kennel_Blandford_Coppi_1989,DELMONT_KEPPENS_2011,
Goedbloed_Keppens_Poedts_2019}. Equation~
\eqref{eq:shock_tangential_invariant_Q} translates this characteristic
ordering into a relation between the corresponding $Q$-branch
combinations.

The distinction entering the invariant is whether the normal Alfv\'en
Mach number remains above unity, remains below unity, or crosses unity.
These cases correspond, respectively, to an increase, a decrease, or a
reversal of the $\alpha$-normalised tangential branch difference between
the upstream and downstream states.

\subsubsection{Fast shocks}
\label{sec:fast_shocks}

For a fast shock, the normal Alfv\'en Mach number remains larger than
unity on both sides of the discontinuity. In the non-limiting case,
\begin{equation}
    M_1^2
    >
    M_2^2
    >
    1.
    \label{eq:fast_M_ordering}
\end{equation}
Since $M^2-1$ is positive and decreases across the shock,
Eq.~\eqref{eq:shock_tangential_invariant_Q} implies that the
$\alpha$-normalised tangential branch difference increases in magnitude
downstream. In $Q$-variables, this is expressed as
\begin{equation}
    \left|
        \frac{
            \boldsymbol{Q}_{t2}^{\prime+}
            -
            \boldsymbol{Q}_{t2}^{\prime-}
        }{\alpha_2}
    \right|
    >
    \left|
        \frac{
            \boldsymbol{Q}_{t1}^{\prime+}
            -
            \boldsymbol{Q}_{t1}^{\prime-}
        }{\alpha_1}
    \right|.
    \label{eq:fast_Q_tangential_amplification}
\end{equation}

The normal momentum condition,
Eq.~\eqref{eq:QRH_normal_momentum}, is
\begin{equation}
    \bigg\llbracket
        \rho
        \left(
            Q_n^{\prime+}
            +
            Q_n^{\prime-}
        \right)^2
        +
        4p
        +
        \frac{1}{2\mu\alpha^2}
        \left|
            \boldsymbol{Q}_t^{\prime+}
            -
            \boldsymbol{Q}_t^{\prime-}
        \right|^2
    \bigg\rrbracket
    =
    0.
    \label{eq:fast_normal_momentum_Q}
\end{equation}
The increase in the tangential branch difference is accompanied by changes
in the normal branch sum and thermodynamic state. Thus, its
$\alpha$-normalised magnitude is larger downstream of a fast shock.

In characteristic terms, the compression changes the local fast
characteristics from
\(\boldsymbol{Q}_{0,f,1}^{\pm}\) upstream to
\(\boldsymbol{Q}_{0,f,2}^{\pm}\) downstream, so that their propagation
speed and, in general, direction differ across the shock. The fast shock
therefore connects two background states with different fast-characteristic
trajectories, while the amplification of the normalised tangential branch
difference gives the corresponding $Q$-space signature of this change.

\subsubsection{Slow shocks}
\label{sec:slow_shocks}

For a slow shock, the normal Alfv\'en Mach number remains smaller than
unity on both sides of the discontinuity. In the non-limiting case,
\begin{equation}
    1
    >
    M_1^2
    >
    M_2^2.
    \label{eq:slow_M_ordering}
\end{equation}
In this case, $M^2-1$ is negative on both sides, while its magnitude
increases downstream. Equation~\eqref{eq:shock_tangential_invariant_Q}
therefore implies that the $\alpha$-normalised tangential branch
difference decreases in magnitude:
\begin{equation}
    \left|
        \frac{
            \boldsymbol{Q}_{t2}^{\prime+}
            -
            \boldsymbol{Q}_{t2}^{\prime-}
        }{\alpha_2}
    \right|
    <
    \left|
        \frac{
            \boldsymbol{Q}_{t1}^{\prime+}
            -
            \boldsymbol{Q}_{t1}^{\prime-}
        }{\alpha_1}
    \right|.
    \label{eq:slow_Q_tangential_reduction}
\end{equation}
Thus, the magnitude of the $\alpha$-normalised tangential branch
difference is smaller downstream of a slow shock.

The corresponding slow characteristics likewise change from
\(\boldsymbol{Q}_{0,s,1}^{\pm}\) to
\(\boldsymbol{Q}_{0,s,2}^{\pm}\) as the plasma crosses the shock. The
slow shock therefore changes the propagation speed and, in general, the
direction of the slow characteristics, while the reduction of the
normalised tangential branch difference provides the characteristic
$Q$-space signature of the slow-shock transition.

\subsubsection{Intermediate shocks}
\label{sec:intermediate_shocks}

Intermediate shocks correspond to cases in which the normal Alfv\'en
Mach number crosses unity across the discontinuity. In the simplest
non-limiting ordering,
\begin{equation}
    M_1^2
    >
    1
    >
    M_2^2.
    \label{eq:intermediate_M_crossing}
\end{equation}
Detailed characteristic-region classifications divide the
intermediate-shock family into several transitions
\citep{Kennel_Blandford_Coppi_1989,DELMONT_KEPPENS_2011}. Their common
property relevant to Eq.~\eqref{eq:shock_tangential_invariant_Q} is that
the upstream and downstream states lie on opposite sides of the normal
Alfv\'enic boundary.

The factor $M^2-1$ changes sign across the shock. Since the quantity
\[
    \left(
        M^2
        -
        1
    \right)
    \frac{
        \boldsymbol{Q}_t^{\prime+}
        -
        \boldsymbol{Q}_t^{\prime-}
    }{\alpha}
\]
is continuous by Eq.~\eqref{eq:shock_tangential_invariant_Q}, the
$\alpha$-normalised tangential branch difference must change orientation
across a non-limiting intermediate shock.

Indeed, Eq.~\eqref{eq:shock_tangential_invariant_Q} gives
\begin{equation}
    \left(
        M_1^2
        -
        1
    \right)
    \frac{
        \boldsymbol{Q}_{t1}^{\prime+}
        -
        \boldsymbol{Q}_{t1}^{\prime-}
    }{\alpha_1}
    =
    \left(
        M_2^2
        -
        1
    \right)
    \frac{
        \boldsymbol{Q}_{t2}^{\prime+}
        -
        \boldsymbol{Q}_{t2}^{\prime-}
    }{\alpha_2}.
    \label{eq:intermediate_vector_invariant}
\end{equation}
Because $M_1^2-1>0$ and $M_2^2-1<0$, the two non-zero normalised
tangential branch differences point in opposite directions.
Equivalently,
\begin{equation}
    \left(
        \frac{
            \boldsymbol{Q}_{t1}^{\prime+}
            -
            \boldsymbol{Q}_{t1}^{\prime-}
        }{\alpha_1}
    \right)
    \cdot
    \left(
        \frac{
            \boldsymbol{Q}_{t2}^{\prime+}
            -
            \boldsymbol{Q}_{t2}^{\prime-}
        }{\alpha_2}
    \right)
    <
    0.
    \label{eq:tangential_branch_reversal_dot}
\end{equation}
Thus, a non-limiting intermediate shock is a compressive discontinuity
across which the $\alpha$-normalised tangential branch difference
reverses orientation.

The characteristic interpretation is particularly direct in this case.
Because the shock connects states on opposite sides of the normal
Alfv\'enic boundary, the relation of the flow to the Alfv\'enic
characteristic changes across the discontinuity. In the shock frame, one
of the normal Alfv\'enic characteristic directions therefore changes its
orientation relative to the shock as \(M^2\) crosses unity. The reversal
of the normalised tangential branch difference is the accompanying
$Q$-space signature of this characteristic transition.

The reversal derived above provides the algebraic $Q$-space signature of
an RH-compatible intermediate shock. Its physical selection additionally
depends on characteristic evolutionarity, dissipative structure,
dimensionality, and the formulation of the corresponding Riemann problem
\citep{FALLE_KOMISSAROV_2001,TAKAHASHI_YAMADA_2013,
Takahashi_Yamada_2014,XU_2024}.
\subsection{Geometrical and limiting shock classes}
\label{sec:geometrical_limiting_shocks}

The geometrical and limiting shock classes provide additional constraints
within the fast, slow, and intermediate families: equality of the normal
branch components for perpendicular shocks, equality of the tangential
branch components for parallel shocks, and appearance or disappearance of
the tangential branch difference for switch-on and switch-off shocks.
Through the characteristic interpretation of the background
$Q$-variables, these conditions correspond to special geometries or
limiting configurations of the associated characteristic propagation.

Switch-on and switch-off shocks occupy limiting positions in the MHD shock
solution space and are associated with degeneracies of the characteristic
structure and with non-regular Riemann solutions
\citep{Kennel_Blandford_Coppi_1989,Takahashi_Yamada_2014}. Here,
appearance and disappearance refer to differences between the two
background shock states.

\subsubsection{Perpendicular shocks}
\label{sec:perpendicular_shocks}

For a perpendicular shock, $B_n=0$. According to
Eq.~\eqref{eq:Q_inverse_relations}, this is equivalent to equality of
the two normal branch components:
\begin{equation}
    Q_n^{\prime+}
    =
    Q_n^{\prime-}
    \equiv
    Q_n^{\prime}.
    \label{eq:perpendicular_normal_collapse_Q}
\end{equation}
The common value is the normal shock-frame velocity,
$Q_n'=V_n'$. Thus, the normal component does not distinguish the
$+$ and $-$ branches in this geometry.

The shock may nevertheless be compressive because the common normal
component is finite and remains constrained by mass conservation. With
Eq.~\eqref{eq:perpendicular_normal_collapse_Q}, the mass condition,
Eq.~\eqref{eq:QRH_mass}, becomes
\begin{equation}
    \bigg\llbracket
        \rho Q_n^{\prime}
    \bigg\rrbracket
    =
    0,
    \label{eq:perpendicular_mass_Q}
\end{equation}
where the common factor of two arising from
$Q_n^{\prime+}+Q_n^{\prime-}=2Q_n'$ has been removed. Therefore, the
two normal branch components coincide, while their common value
determines the finite normal mass flux through the shock.

The tangential magnetic-flux condition,
Eq.~\eqref{eq:QRH_tangential_flux}, reduces to
\begin{equation}
    \bigg\llbracket
        Q_n^{\prime}
        \frac{
            \boldsymbol{Q}_t^{\prime+}
            -
            \boldsymbol{Q}_t^{\prime-}
        }{\alpha}
    \bigg\rrbracket
    =
    0.
    \label{eq:perpendicular_tangential_flux_Q}
\end{equation}
Using Eq.~\eqref{eq:perpendicular_mass_Q}, this may be written as
\begin{equation}
    \bigg\llbracket
        \frac{
            \boldsymbol{Q}_t^{\prime+}
            -
            \boldsymbol{Q}_t^{\prime-}
        }{
            \rho\alpha
        }
    \bigg\rrbracket
    =
    0.
    \label{eq:perpendicular_tangential_flux_density_Q}
\end{equation}
The tangential momentum condition,
Eq.~\eqref{eq:QRH_tangential_momentum}, gives
\begin{equation}
    \bigg\llbracket
        \boldsymbol{Q}_t^{\prime+}
        +
        \boldsymbol{Q}_t^{\prime-}
    \bigg\rrbracket
    =
    0.
    \label{eq:perpendicular_tangential_sum_Q}
\end{equation}
Finally, the normal momentum balance,
Eq.~\eqref{eq:QRH_normal_momentum}, becomes
\begin{equation}
    \bigg\llbracket
        4\rho
        \left(
            Q_n^{\prime}
        \right)^2
        +
        4p
        +
        \frac{1}{2\mu\alpha^2}
        \left|
            \boldsymbol{Q}_t^{\prime+}
            -
            \boldsymbol{Q}_t^{\prime-}
        \right|^2
    \bigg\rrbracket
    =
    0.
    \label{eq:perpendicular_normal_momentum_Q}
\end{equation}

Collecting these relations, a perpendicular shock is described by
\begin{align}
    &\llbracket \rho \rrbracket
    \neq
    0,
    \qquad
    Q_n^{\prime+}
    =
    Q_n^{\prime-}
    \equiv
    Q_n^{\prime},
    \nonumber\\[4pt]
    &\bigg\llbracket
        \rho Q_n^{\prime}
    \bigg\rrbracket
    =
    0,
    \nonumber\\[4pt]
    &\bigg\llbracket
        \frac{
            \boldsymbol{Q}_t^{\prime+}
            -
            \boldsymbol{Q}_t^{\prime-}
        }{
            \rho\alpha
        }
    \bigg\rrbracket
    =
    0,
    \nonumber\\[4pt]
    &\bigg\llbracket
        \boldsymbol{Q}_t^{\prime+}
        +
        \boldsymbol{Q}_t^{\prime-}
    \bigg\rrbracket
    =
    0,
    \nonumber\\[4pt]
    &\bigg\llbracket
        4\rho
        \left(
            Q_n^{\prime}
        \right)^2
        +
        4p
        +
        \frac{1}{2\mu\alpha^2}
        \left|
            \boldsymbol{Q}_t^{\prime+}
            -
            \boldsymbol{Q}_t^{\prime-}
        \right|^2
    \bigg\rrbracket
    =
    0.
    \label{eq:perpendicular_summary_Q}
\end{align}

For \(B_n=0\), the normal Alfv\'en Mach number in
Eq.~\eqref{eq:shock_Mach_Q} becomes singular because its $Q$-space
denominator vanishes. The perpendicular shock is therefore classified
directly through its geometrical conditions: the two normal branch
components coincide, while the tangential branch difference may remain
finite and satisfies
Eq.~\eqref{eq:perpendicular_tangential_flux_density_Q}.

In characteristic terms, the perpendicular geometry removes the
distinction between the $+$ and $-$ branches in their normal
$Q$-component: both have the same normal projection in the shock frame.
The characteristic modification across the shock is therefore not
encoded in a separation of the two normal branch components, but in the
change of their common normal component together with the tangential
branch structure. Compression changes the background state and hence the
corresponding characteristic propagation, while the condition \(B_n=0\)
imposes this geometrical degeneracy of the normal $Q$-branch projection.

\subsubsection{Parallel shocks}
\label{sec:parallel_shocks}

For a parallel shock, $\boldsymbol{B}_t=\boldsymbol{0}$. According to
Eq.~\eqref{eq:Q_inverse_relations}, this is equivalent to equality of
the two tangential branch components:
\begin{equation}
    \boldsymbol{Q}_t^{\prime+}
    =
    \boldsymbol{Q}_t^{\prime-}
    \equiv
    \boldsymbol{Q}_t^{\prime}.
    \label{eq:parallel_tangential_collapse_Q}
\end{equation}
Their common value is the tangential shock-frame velocity,
$\boldsymbol{Q}_t'=\boldsymbol{V}_t'$. Thus, the tangential component
does not distinguish the $+$ and $-$ branches in this geometry.

The mass and normal magnetic-flux conditions,
Eqs.~\eqref{eq:QRH_mass} and~\eqref{eq:QRH_normal_flux}, are
\begin{align}
    &\bigg\llbracket
        \rho
        \left(
            Q_n^{\prime+}
            +
            Q_n^{\prime-}
        \right)
    \bigg\rrbracket
    =
    0,
    \label{eq:parallel_mass_Q}
    \\[4pt]
    &\bigg\llbracket
        \frac{
            Q_n^{\prime+}
            -
            Q_n^{\prime-}
        }{\alpha}
    \bigg\rrbracket
    =
    0.
    \label{eq:parallel_normal_flux_Q}
\end{align}
The tangential magnetic-flux and tangential momentum conditions,
Eqs.~\eqref{eq:QRH_tangential_flux}
and~\eqref{eq:QRH_tangential_momentum}, imply continuity of the common
tangential component:
\begin{equation}
    \bigg\llbracket
        \boldsymbol{Q}_t^{\prime}
    \bigg\rrbracket
    =
    0.
    \label{eq:parallel_tangential_continuity_Q}
\end{equation}
The normal momentum condition,
Eq.~\eqref{eq:QRH_normal_momentum}, reduces to
\begin{equation}
    \bigg\llbracket
        \rho
        \left(
            Q_n^{\prime+}
            +
            Q_n^{\prime-}
        \right)^2
        +
        4p
    \bigg\rrbracket
    =
    0.
    \label{eq:parallel_normal_momentum_Q}
\end{equation}

Collecting these relations, a parallel shock is described by
\begin{align}
    &\llbracket \rho \rrbracket
    \neq
    0,
    \qquad
    \bigg\llbracket
        Q_n^{\prime+}
        +
        Q_n^{\prime-}
    \bigg\rrbracket
    \neq
    0,
    \nonumber\\[4pt]
    &\boldsymbol{Q}_t^{\prime+}
    =
    \boldsymbol{Q}_t^{\prime-}
    \equiv
    \boldsymbol{Q}_t^{\prime},
    \nonumber\\[4pt]
    &\bigg\llbracket
        \rho
        \left(
            Q_n^{\prime+}
            +
            Q_n^{\prime-}
        \right)
    \bigg\rrbracket
    =
    0,
    \nonumber\\[4pt]
    &\bigg\llbracket
        \frac{
            Q_n^{\prime+}
            -
            Q_n^{\prime-}
        }{\alpha}
    \bigg\rrbracket
    =
    0,
    \nonumber\\[4pt]
    &\bigg\llbracket
        \boldsymbol{Q}_t^{\prime}
    \bigg\rrbracket
    =
    0,
    \nonumber\\[4pt]
    &\bigg\llbracket
        \rho
        \left(
            Q_n^{\prime+}
            +
            Q_n^{\prime-}
        \right)^2
        +
        4p
    \bigg\rrbracket
    =
    0.
    \label{eq:parallel_summary_Q}
\end{align}
Thus, a parallel shock is a compressive geometrical limit in which the two
tangential branch components coincide and remain continuous. Its magnetic
branch structure is contained in the normal branch difference, while the
normal branch sum and thermodynamic variables satisfy the compressive RH
conditions.

In the characteristic picture, the parallel geometry removes the
$+$/$-$ branch separation from the tangential direction, so that the
wave-aligned distinction between the two branches is carried by their
normal components. Compression changes these normal components and hence
the characteristic propagation speed across the shock. Since the common
tangential component is continuous, any change in characteristic
orientation arises from the change in the normal propagation relative to
this unchanged tangential component.

\subsubsection{Switch-on shocks}
\label{sec:switch_on_shocks}

A switch-on shock is a limiting shock for which the tangential branch
difference vanishes upstream and is finite downstream. In $Q$-variables,
this is expressed as
\begin{equation}
    \boldsymbol{Q}_{t1}^{\prime+}
    =
    \boldsymbol{Q}_{t1}^{\prime-},
    \qquad
    \boldsymbol{Q}_{t2}^{\prime+}
    \neq
    \boldsymbol{Q}_{t2}^{\prime-}.
    \label{eq:switch_on_Q}
\end{equation}
The tangential invariant
\eqref{eq:shock_tangential_invariant_Q} shows why this is a limiting
case. Since the upstream tangential branch difference vanishes, a
finite downstream difference requires
\begin{equation}
    M_2^2
    =
    1.
    \label{eq:switch_on_M_condition}
\end{equation}
Thus, the downstream state lies on the Alfv\'enic boundary. In
$Q$-space, the tangential branch difference is absent in the upstream
state but present in the downstream state.

The characteristic interpretation reflects this limiting behaviour.
Upstream, the tangential $+$ and $-$ branches coincide, whereas
downstream they separate and the state terminates on the Alfv\'enic
characteristic boundary, \(M_2^2=1\). A switch-on shock therefore
connects a state with no tangential branch separation to a state with a
finite tangential characteristic structure, with the downstream
characteristic configuration constrained to the Alfv\'enic limit.

\subsubsection{Switch-off shocks}
\label{sec:switch_off_shocks}

A switch-off shock is a limiting shock for which the tangential branch
difference is finite upstream and vanishes downstream. In $Q$-variables,
this is expressed as
\begin{equation}
    \boldsymbol{Q}_{t1}^{\prime+}
    \neq
    \boldsymbol{Q}_{t1}^{\prime-},
    \qquad
    \boldsymbol{Q}_{t2}^{\prime+}
    =
    \boldsymbol{Q}_{t2}^{\prime-}.
    \label{eq:switch_off_Q}
\end{equation}
Again, the tangential invariant
\eqref{eq:shock_tangential_invariant_Q} shows that this is a limiting
case. Since the downstream tangential branch difference vanishes while
the upstream difference is finite, one must have
\begin{equation}
    M_1^2
    =
    1.
    \label{eq:switch_off_M_condition}
\end{equation}
Thus, the upstream state lies on the Alfv\'enic boundary. In $Q$-space,
the tangential branch difference is present in the upstream state but
absent in the downstream state.

The characteristic interpretation is the reverse of the switch-on case.
The upstream state lies on the Alfv\'enic characteristic boundary and
contains a finite tangential branch separation, whereas the downstream
$+$ and $-$ tangential branches coincide. A switch-off shock therefore
removes this tangential characteristic structure as the plasma passes
through the shock, with the upstream state providing the limiting
Alfv\'enic configuration.

Table~\ref{tab:Q_classification} collects the resulting $Q$-space
signatures together with their characteristic interpretation.

\section{Classification table}
\label{sec:classification_table}

Table~\ref{tab:Q_classification} collects the branch relations derived
above together with their characteristic interpretation. It links the
classical characteristic and geometrical organisation of ideal-MHD
discontinuities to their wave-aligned $Q$-space signatures.

\begin{table*}[p]
\caption{
$Q$-space classification of ideal-MHD discontinuities and their
characteristic interpretation. All $Q$-variables are evaluated in the
discontinuity frame, and subscripts 1 and 2 denote the states on its two
sides. For contact and tangential discontinuities, a density jump is
permitted but is not a defining condition. The fast, slow, and
intermediate entries show the non-limiting cases.
}
\label{tab:Q_classification}

\centering
\scriptsize
\renewcommand{\arraystretch}{1.18}
\setlength{\tabcolsep}{2.5pt}


\begin{tabular}{@{}llll@{}}
\toprule

\parbox[t]{0.115\textwidth}{\raggedright
Discontinuity
}
&
\parbox[t]{0.205\textwidth}{\raggedright
Normal / Mach signature
}
&
\parbox[t]{0.315\textwidth}{\raggedright
Tangential signature
}
&
\parbox[t]{0.305\textwidth}{\raggedright
Characteristic interpretation
}
\\

\midrule

\parbox[t]{0.115\textwidth}{\raggedright
Contact
}
&
\parbox[t]{0.205\textwidth}{\raggedright
\(\displaystyle
Q_n^{\prime+}
=
-Q_n^{\prime-}
\).
\par\smallskip
\(\displaystyle
\frac{Q_n^{\prime\pm}}{\alpha}
\neq
0
\).
\par\smallskip
\(\displaystyle
\left\llbracket
\frac{Q_n^{\prime\pm}}{\alpha}
\right\rrbracket
=
0
\).
}
&
\parbox[t]{0.315\textwidth}{\raggedright
\(\displaystyle
\left\llbracket
\boldsymbol Q_t^{\prime+}
+
\boldsymbol Q_t^{\prime-}
\right\rrbracket
=
0
\).
\par\smallskip
\(\displaystyle
\left\llbracket
\frac{
\boldsymbol Q_t^{\prime+}
-
\boldsymbol Q_t^{\prime-}
}{\alpha}
\right\rrbracket
=
0
\).
\par\smallskip
\(\displaystyle
\llbracket p\rrbracket
=
0
\).
}
&
\parbox[t]{0.305\textwidth}{\raggedright
Finite normal propagation; a density jump may change characteristic
speed and direction.
}
\\

\midrule

\parbox[t]{0.115\textwidth}{\raggedright
Tangential
}
&
\parbox[t]{0.205\textwidth}{\raggedright
\(\displaystyle
Q_n^{\prime+}
=
Q_n^{\prime-}
=
0
\).
}
&
\parbox[t]{0.315\textwidth}{\raggedright
\(\displaystyle
\left\llbracket
4p
+
\frac{
\left|
\boldsymbol Q_t^{\prime+}
-
\boldsymbol Q_t^{\prime-}
\right|^2
}{
2\mu\alpha^2
}
\right\rrbracket
=
0
\).
}
&
\parbox[t]{0.305\textwidth}{\raggedright
No normal characteristic propagation; structure is tangential.
}
\\

\midrule

\parbox[t]{0.115\textwidth}{\raggedright
Rotational
}
&
\parbox[t]{0.205\textwidth}{\raggedright
\(\displaystyle
\llbracket\rho\rrbracket=0,
\qquad
\llbracket p\rrbracket=0
\).
\par\smallskip
\(\displaystyle
Q_n^{\prime+}
+
Q_n^{\prime-}
\neq0
\).
\par\smallskip
\(\displaystyle
\left\llbracket
Q_n^{\prime+}
+
Q_n^{\prime-}
\right\rrbracket
=
0
\).
\par\smallskip
\(\displaystyle
\left\llbracket
\frac{
Q_n^{\prime+}
-
Q_n^{\prime-}
}{\alpha}
\right\rrbracket
=
0
\).
}
&
\parbox[t]{0.315\textwidth}{\raggedright
\(\displaystyle
M^2=1
\).
\par\smallskip
\(\displaystyle
\left\llbracket
\boldsymbol Q_t^{\prime+}
+
\boldsymbol Q_t^{\prime-}
\right\rrbracket
=
\pm
\frac{1}{\sqrt{\mu\rho}}
\left\llbracket
\frac{
\boldsymbol Q_t^{\prime+}
-
\boldsymbol Q_t^{\prime-}
}{\alpha}
\right\rrbracket
\).
\par\smallskip
\(\displaystyle
\left\llbracket
\frac{
\left|
\boldsymbol Q_t^{\prime+}
-
\boldsymbol Q_t^{\prime-}
\right|^2
}{\alpha^2}
\right\rrbracket
=
0
\).
}
&
\parbox[t]{0.305\textwidth}{\raggedright
Alfv\'enic characteristic; tangential rotation bends the characteristic.
}
\\

\bottomrule
\end{tabular}


\vspace{4pt}

\begin{minipage}{0.958\textwidth}
\raggedright
\textit{Common compressive-shock conditions:}

\smallskip

\(\displaystyle
\llbracket\rho\rrbracket\neq0,
\qquad
Q_n^{\prime+}+Q_n^{\prime-}\neq0,
\qquad
\left\llbracket
Q_n^{\prime+}+Q_n^{\prime-}
\right\rrbracket\neq0,
\qquad
\left\llbracket
\rho
\left(
Q_n^{\prime+}
+
Q_n^{\prime-}
\right)
\right\rrbracket
=
0,
\qquad
\left\llbracket
\frac{
Q_n^{\prime+}
-
Q_n^{\prime-}
}{\alpha}
\right\rrbracket
=
0.
\)

\end{minipage}

\vspace{4pt}


\begin{tabular}{@{}llll@{}}
\toprule

\parbox[t]{0.115\textwidth}{\raggedright
Fast shock
}
&
\parbox[t]{0.205\textwidth}{\raggedright
\(\displaystyle
M_1^2
>
M_2^2
>
1
\).
}
&
\parbox[t]{0.315\textwidth}{\raggedright
\(\displaystyle
\left|
\frac{
\boldsymbol Q_{t2}^{\prime+}
-
\boldsymbol Q_{t2}^{\prime-}
}{\alpha_2}
\right|
>
\left|
\frac{
\boldsymbol Q_{t1}^{\prime+}
-
\boldsymbol Q_{t1}^{\prime-}
}{\alpha_1}
\right|
\).
}
&
\parbox[t]{0.305\textwidth}{\raggedright
Fast-characteristic speed/direction changes across the shock.
}
\\

\midrule

\parbox[t]{0.115\textwidth}{\raggedright
Slow shock
}
&
\parbox[t]{0.205\textwidth}{\raggedright
\(\displaystyle
1
>
M_1^2
>
M_2^2
\).
}
&
\parbox[t]{0.315\textwidth}{\raggedright
\(\displaystyle
\left|
\frac{
\boldsymbol Q_{t2}^{\prime+}
-
\boldsymbol Q_{t2}^{\prime-}
}{\alpha_2}
\right|
<
\left|
\frac{
\boldsymbol Q_{t1}^{\prime+}
-
\boldsymbol Q_{t1}^{\prime-}
}{\alpha_1}
\right|
\).
}
&
\parbox[t]{0.305\textwidth}{\raggedright
Slow-characteristic speed/direction changes across the shock.
}
\\

\midrule

\parbox[t]{0.115\textwidth}{\raggedright
Intermediate shock
}
&
\parbox[t]{0.205\textwidth}{\raggedright
\(\displaystyle
M_1^2
>
1
>
M_2^2
\).
}
&
\parbox[t]{0.315\textwidth}{\raggedright
\(\displaystyle
\left(
\frac{
\boldsymbol Q_{t1}^{\prime+}
-
\boldsymbol Q_{t1}^{\prime-}
}{\alpha_1}
\right)
\!\cdot\!
\left(
\frac{
\boldsymbol Q_{t2}^{\prime+}
-
\boldsymbol Q_{t2}^{\prime-}
}{\alpha_2}
\right)
<
0
\).
}
&
\parbox[t]{0.305\textwidth}{\raggedright
Crosses the Alfv\'enic characteristic boundary; orientation reverses.
}
\\

\midrule

\parbox[t]{0.115\textwidth}{\raggedright
Perpendicular shock
}
&
\parbox[t]{0.205\textwidth}{\raggedright
\(\displaystyle
Q_n^{\prime+}
=
Q_n^{\prime-}
\equiv
Q_n^{\prime}
\).
\par\smallskip
\(\displaystyle
\left\llbracket
\rho Q_n^{\prime}
\right\rrbracket
=
0
\).
\par\smallskip
\(M^2\) in Eq.~\eqref{eq:shock_Mach_Q} is not defined.
}
&
\parbox[t]{0.315\textwidth}{\raggedright
\(\displaystyle
\left\llbracket
\frac{
\boldsymbol Q_t^{\prime+}
-
\boldsymbol Q_t^{\prime-}
}{\rho\alpha}
\right\rrbracket
=
0
\).
\par\smallskip
\(\displaystyle
\left\llbracket
\boldsymbol Q_t^{\prime+}
+
\boldsymbol Q_t^{\prime-}
\right\rrbracket
=
0
\).
}
&
\parbox[t]{0.305\textwidth}{\raggedright
Normal \(+\) and \(-\) characteristic projections coincide.
}
\\

\midrule

\parbox[t]{0.115\textwidth}{\raggedright
Parallel shock
}
&
\parbox[t]{0.205\textwidth}{\raggedright
Common compressive-shock conditions.
}
&
\parbox[t]{0.315\textwidth}{\raggedright
\(\displaystyle
\boldsymbol Q_t^{\prime+}
=
\boldsymbol Q_t^{\prime-}
\equiv
\boldsymbol Q_t^{\prime}
\).
\par\smallskip
\(\displaystyle
\left\llbracket
\boldsymbol Q_t^{\prime}
\right\rrbracket
=
0
\).
}
&
\parbox[t]{0.305\textwidth}{\raggedright
Tangential \(+\) and \(-\) characteristic projections coincide.
}
\\

\midrule

\parbox[t]{0.115\textwidth}{\raggedright
Switch-on shock
}
&
\parbox[t]{0.205\textwidth}{\raggedright
\(\displaystyle
M_2^2
=
1
\).
}
&
\parbox[t]{0.315\textwidth}{\raggedright
\(\displaystyle
\boldsymbol Q_{t1}^{\prime+}
=
\boldsymbol Q_{t1}^{\prime-}
\).
\par\smallskip
\(\displaystyle
\boldsymbol Q_{t2}^{\prime+}
\neq
\boldsymbol Q_{t2}^{\prime-}
\).
}
&
\parbox[t]{0.305\textwidth}{\raggedright
Downstream state lies on the Alfv\'enic characteristic boundary.
}
\\

\midrule

\parbox[t]{0.115\textwidth}{\raggedright
Switch-off shock
}
&
\parbox[t]{0.205\textwidth}{\raggedright
\(\displaystyle
M_1^2
=
1
\).
}
&
\parbox[t]{0.315\textwidth}{\raggedright
\(\displaystyle
\boldsymbol Q_{t1}^{\prime+}
\neq
\boldsymbol Q_{t1}^{\prime-}
\).
\par\smallskip
\(\displaystyle
\boldsymbol Q_{t2}^{\prime+}
=
\boldsymbol Q_{t2}^{\prime-}
\).
}
&
\parbox[t]{0.305\textwidth}{\raggedright
Upstream state lies on the Alfv\'enic characteristic boundary.
}
\\

\bottomrule
\end{tabular}

\end{table*}

\clearpage

\section{Discussion}
\label{sec:discussion}

The classification developed here provides a wave-branch reading of the
standard ideal-MHD discontinuity catalogue. The $Q$-variable formulation
retains the RH content while reorganising the velocity-magnetic-field
state according to the discontinuity-frame branches
\(\boldsymbol Q^{\prime+}\) and \(\boldsymbol Q^{\prime-}\).

\subsection{Physical meaning of the
\texorpdfstring{$Q$}{Q}-space classification}

The value of the $Q$-variable classification is that it reorganises
information that is combined in the primitive variables. In the classical
formulation, velocity and magnetic field appear as separate plasma
variables. In $Q$-space, the same velocity-magnetic-field state is written
in terms of two wave-aligned branches. For homogeneous fast, Alfv\'en, and
slow waves, the branch parameter selects a representation associated with
the corresponding MHD characteristic, whereas for structured waves such
as kink modes it selects the relevant phase relation. The $Q$-formalism
therefore connects the characteristic structure organising the classical
discontinuity families with a broader wave-aligned description.

The RH conditions determine which branch combinations are continuous,
coupled, equal, or vanishing across the interface. The resulting relations
describe the finite background states on its two sides and provide the
matching structure about which a specific wave interaction can be
formulated.

\subsection{Characteristic interpretation}
\label{sec:characteristic_interpretation}

For a branch-specific choice of \(\alpha_m\), the background variables
\(\boldsymbol Q_{0,m}^{\pm}\) represent the corresponding fast,
Alfv\'en, or slow characteristics. As introduced in
Section~\ref{sec:introduction}, their normal projections satisfy
\[
    Q_{0,m,n}^{\pm}
    =
    \lambda_m^{\pm},
\]
so that the $Q$-variables provide a direct representation of the
propagation of the corresponding characteristic. A discontinuity
separates two background states and therefore, in general, two different
sets of characteristic vectors,
\[
    \boldsymbol Q_{0,m,1}^{\pm}
    \longrightarrow
    \boldsymbol Q_{0,m,2}^{\pm}.
\]
The $Q$-RH conditions determine how these two characteristic
configurations are matched across the interface.

This gives the $Q$-space classification a geometrical interpretation.
A change in the normal projection of a background branch corresponds to
a change in the characteristic propagation speed relative to the
interface. More generally, if the normal and tangential components change
such that
\(\boldsymbol Q_{0,m,1}^{\pm}\) and
\(\boldsymbol Q_{0,m,2}^{\pm}\) are not parallel, the direction of
characteristic propagation changes across the discontinuity. In this
sense, the characteristic is bent by the discontinuity. The jump
conditions therefore encode the corresponding change in characteristic speed and direction.

The different discontinuity families realise this characteristic
modification in different ways. Material interfaces constrain the normal
characteristic propagation through their vanishing or finite normal
$Q$-components. Rotational discontinuities provide the clearest
Alfv\'enic example: the condition \(M^2=1\) places the discontinuity on
the normal Alfv\'enic characteristic, while the tangential structure may
rotate across the interface. Compressive shocks connect different
background characteristic configurations through the change of plasma
state; intermediate shocks additionally cross the normal Alfv\'enic
boundary. Perpendicular and parallel shocks impose geometrical
degeneracies of the normal or tangential branch projections, while
switch-on and switch-off shocks terminate on limiting Alfv\'enic states.

\subsection{Diagnostic use}
\label{sec:diagnostic_use}

The classification table suggests a diagnostic procedure for
simulations and observations. Once the discontinuity normal,
propagation speed, upstream and downstream states, and branch parameter
$\alpha$ are specified, the discontinuity-frame $Q$-branches can be
constructed on both sides and their algebraic signatures compared with
those in Table~\ref{tab:Q_classification}.

In spacecraft applications, the required discontinuity normal and
propagation speed are generally inferred using methods such as magnetic
or velocity coplanarity, minimum-variance analysis, de
Hoffmann-Teller analysis, timing methods, or multi-spacecraft
measurements \citep{Knetter_2004,Neugebauer_2006}. The resulting
classification can be sensitive to uncertainties in the normal and to
the thresholds used to distinguish, for example, tangential from
rotational discontinuities. The exact $Q$-space conditions derived here
should therefore be interpreted as theoretical signatures from which
uncertainty-aware diagnostic residuals may subsequently be constructed.

Complementary discontinuity diagnostics have recently been constructed
from the velocity jump and the magnetic-field jump expressed in
Alfv\'en units, including residual-energy and cross-helicity signatures
for fast, slow, contact, tangential, and rotational discontinuities
\citep{GOOD_2025}. The present $Q$-space classification complements
these diagnostics by retaining the signed vectorial branch information,
separating the normal and tangential components, and allowing a general
branch parameter $\alpha$. A quantitative comparison between these
diagnostic descriptions is left for future work.

The density and normal branch components first distinguish between
no-mass-flux interfaces, non-compressive finite-flux discontinuities,
and compressive shocks. The normal magnetic-field condition then
separates contact from tangential discontinuities, while the tangential
relations and Mach-number conditions provide the signatures of
rotational discontinuities and the different shock classes.

In practical applications, the diagnostic quality will depend on how
well the discontinuity normal, propagation speed, upstream and
downstream states, and branch parameter $\alpha$ are determined. These
quantities may be uncertain in spacecraft measurements or in
finite-thickness structures in simulations. Nevertheless, once they are
specified, the $Q$-variable classification provides an additional
diagnostic layer by showing how the wave-aligned representation of the
state is constrained across the discontinuity.

\subsection{Role of the branch parameter
\texorpdfstring{\(\alpha\)}{alpha}}
\label{sec:role_of_alpha}

A central feature of the $Q$-variable description is the branch
parameter $\alpha$. The classification is written in terms of
$\alpha$-normalised branch differences because these are the
combinations that appear naturally in the $Q$-RH conditions. The
parameter $\alpha$ should therefore not be viewed only as a scaling of
the magnetic field: it specifies the wave-aligned representation used
to express the discontinuity states.

For the Alfv\'enic branch,
$\alpha=1/\sqrt{\mu\rho}$, and the $Q$-variables reduce to the
Els\"asser variables. In this limit, the Alfv\'enic character of
rotational discontinuities is represented particularly transparently
\citep{krupka2026rankine}. For fast, slow, kink, or other
structured-plasma branches, $\alpha$ must instead be determined from
the corresponding wave relation and local plasma state. Applying the
same branch prescription on the two sides does not generally imply
$\alpha_1=\alpha_2$, because the local plasma parameters may change
across the discontinuity.

The classification table therefore does not assume a single universal
value of $\alpha$. It gives the discontinuity constraints for a chosen
wave-aligned representation, allowing the same plasma jump to be
expressed using the $\alpha$ appropriate to different selected wave
branches. This also determines which background characteristic is being
represented when the characteristic interpretation is applied.

\subsection{Connection to wave-discontinuity interactions}
\label{sec:relation_to_wave_interactions}

The present paper classifies discontinuities; it does not solve a
scattering problem. The RH conditions used here relate the nonlinear
background states on the two sides of a discontinuity. They do not, by
themselves, determine incident, reflected, or transmitted wave
amplitudes. However, a wave-discontinuity interaction theory needs
precisely the structural information obtained here. Before one can ask
how a selected wave interacts with an interface, one must know what
branch-level constraints that interface imposes.

Similarly, the terms amplification, reduction, reversal, appearance,
and disappearance used in the shock classification describe how a
particular branch combination differs between the upstream and
downstream background states. They do not describe the amplification,
conversion, creation, or absorption of a propagating perturbation. The
characteristic bending or change in propagation speed discussed above
likewise refers to the characteristics of the two background states,
rather than to the scattering of a finite-amplitude incident wave.

A wave-discontinuity interaction theory nevertheless requires the
background structure obtained here. A future calculation can select an
incoming Alfv\'en, fast, slow, or kink perturbation and linearise or
otherwise perturb the jump conditions about one of the classified
discontinuity states. Together with the appropriate dispersion
relations, interface kinematics, and boundary conditions, this would
determine the reflected, transmitted, and mode-converted amplitudes.

This also clarifies why the classification precedes the interaction
problem. A statement such as ``a fast wave meets a slow shock'' combines
two separate pieces of information: the branch content of the incoming
disturbance and the background signature of the discontinuity. The
present paper determines the second part. Future work can combine it
with a specified incoming disturbance to formulate and solve the
corresponding coupling or scattering problem.

\subsection{Limitations and outlook}
\label{sec:limitations_outlook}

The classification applies to the asymptotic states of infinitesimally
thin ideal-MHD discontinuities. Kinetic-scale current sheets may possess
finite thickness and internal density and temperature structure
\citep{Neukirch_2020}; the corresponding non-ideal layer lies between
the asymptotic states classified here.

A consistent branch prescription for \(\alpha\) is applied on both
sides. For magneto-acoustic and structured-plasma waves, its local values
follow from the plasma state and the relevant wave relation. Their
influence on quantitative event identification forms part of the
subsequent diagnostic development.

The RH relations provide the algebraic foundation for the
classification. For intermediate and other non-regular shocks,
characteristic evolutionarity, dissipative structure, stability, and
uniqueness of the Riemann solution supply additional physical selection
criteria
\citep{FALLE_KOMISSAROV_2001,TAKAHASHI_YAMADA_2013,
Takahashi_Yamada_2014,XU_2024}.

The resulting interface map provides the starting point for calculations
of reflection, transmission, coupling, and mode conversion for specific
MHD wave branches.

\section{Conclusions}
\label{sec:conclusions}

We have constructed a wave-branch classification of ideal-MHD
discontinuities in $Q$-variables. Starting from the $Q$-variable RH
conditions, we determined how the discontinuity-frame branches
\(\boldsymbol{Q}^{\prime+}\) and \(\boldsymbol{Q}^{\prime-}\), their
sums, and their \(\alpha\)-normalised differences are constrained across
the standard discontinuity families.

The classical characteristic structure organises the discontinuities,
while the branch parameter \(\alpha\) selects the wave representation in
which their interface constraints are expressed. For homogeneous ideal
MHD, the corresponding background $Q$-variables represent the fast,
Alfv\'en, and slow characteristics, while for structured plasmas the same
formalism extends to global modes such as kink waves.

This characteristic interpretation gives the $Q$-space jump relations a
direct geometrical meaning. A discontinuity generally connects two
different background characteristic configurations. Changes in the
normal branch components correspond to changes in characteristic
propagation relative to the interface, while changes in the orientation
of the corresponding $Q$-vectors describe bending of the
characteristics across the discontinuity. The $Q$-RH conditions therefore
encode how characteristic propagation is matched between the two plasma
states.

Material interfaces are distinguished by their normal branch content,
rotational discontinuities by an Alfv\'enic tangential relation, and
compressive shocks by the behaviour of the density-weighted normal branch
sum and the $\alpha$-normalised tangential branch difference. Rotational
discontinuities lie on the Alfv\'enic characteristic and allow the
tangential structure to rotate, while fast and slow shocks modify the
corresponding characteristic configurations across the compressed states.
Intermediate shocks cross the normal Alfv\'enic boundary, and
perpendicular, parallel, switch-on, and switch-off shocks arise as
geometrical or limiting cases.

The resulting map therefore provides both a \(Q\)-space classification
of the ideal-MHD discontinuity catalogue and a characteristic
interpretation of how each interface modifies the propagation of the
background state. It also establishes the background interface structure
required for wave-discontinuity interaction problems. In this sense,
the relations derived here may be viewed as the zeroth-order counterpart
of a Snell-type wave-interface problem: linearisation about these
matched background states provides the next step toward determining
reflection, transmission, coupling, and mode conversion for specified
incident wave branches in the \(Q\)-variable framework.

\section*{Funding}
T.V.D. was supported by a Senior Research Project (G088021N) of the FWO
Vlaanderen. Furthermore, TVD received financial support from the Flemish
Government under the long-term structural Methusalem funding program,
project SOUL: Stellar evolution in full glory, grant METH/24/012 at KU
Leuven. The research that led to these results was subsidised by the
Belgian Federal Science Policy Office through the contract
B2/223/P1/CLOSE-UP. It is also part of the DynaSun project and has thus
received funding under the Horizon Europe programme of the European
Union under grant agreement (no. 101131534). Views and opinions
expressed are however those of the author(s) only and do not necessarily
reflect those of the European Union and therefore the European Union
cannot be held responsible for them.
\section*{Declaration of interests}

The authors report no conflict of interest.

\bibliographystyle{apsrev4-1}
\bibliography{bibliography}
\end{document}